\documentclass[1p,11pt]{elsarticle}

\usepackage{lineno,hyperref}
\modulolinenumbers[5]

\usepackage{amsmath}
\usepackage{amssymb}
\usepackage{xspace} \usepackage{ifthen}
\usepackage{array}
\usepackage{xcolor}
\usepackage{url}

\newboolean{figurelist}
\setboolean{figurelist}{true}

\newcommand	{\incfig}	[3]	{\ifthenelse{\boolean{figurelist}}
		{\immediate\write\outstream{fig-#1.pdf}}
	{}
\begin{figure}[!t]
    \includegraphics
				[#2]	{fig-#1}
    \caption{#3}
    \label{fig:#1}
	\end{figure}
	}

\newcommand {\doctable}
	[5] {
	\begin{table}[!t]
\scriptsize
	\caption{#2}
	\label{tbl:#1}
	\centering
	\begin{tabular}{#3}
#4
	\end{tabular}
	\end{table}
	}

\newenvironment
	{example}
	{
	\begin{description}
	\item [Example:]
	}
	{\end{description}}

\newenvironment{makefigurelist}
	{
\ifthenelse{\boolean{figurelist}}
		{
		\newwrite\outstream
		\immediate\openout\outstream=figure_list
}
		{} }
	{
\ifthenelse{\boolean{figurelist}}
		{
		\immediate\closeout\outstream
		}
		{}
	}

\newcommand{\eqnref}[1] {Eq.\eqref{eq:#1}}
\newcommand{\secref}[1] {\S\ref{sec:#1}}
\newcommand{\tblref}[1] {Tbl.\ref{tbl:#1}}
\newcommand{\figref}[1] {Fig.\ref{fig:#1}}

\newcommand{\ile}[1]{\mbox{$#1$}}

\newcommand{\sss}[2]{#1^{}_{\!_{#2}}}
\newcommand{\ssss}[3]{#1^{#3}_{\!_{#2}}}

\newcommand{\distanceTarget}{\sss{D}{\text{star}}}
\newcommand{\distanceTargetSquared}{\ssss{D}{\text{star}}{2}}

\newcommand{\timeCoord}{t}
\newcommand{\timeTransmission}{\sss{T}{\text{down}}}
\newcommand{\speedProbe}[1]{\sss{u}{\,#1}}
\newcommand{\speedLight}{c}
\newcommand{\timeLaunch}{\sss{T}{\text{launch}}}
\newcommand{\timeLaunchSquared}{T^{\,2}_{\!_{\,\text{launch}}}}

\newcommand{\energyLaunch}{\sss{\mathcal E}{\text{launch}}}

\newcommand{\rateData}[1]{\sss{\mathcal R}{\,#1}}
\newcommand{\dataVolume}{\sss{\mathcal V}{\text{data}}}
\newcommand{\dataVolumeNorm}{\dataVolume/\rateData{0}}
\newcommand{\dataLatency}{\sss{T}{\text{latency}}}

\newcommand{\mass}[1]{\sss{m}{\,#1}}
\newcommand{\massE}[2]{\ssss{m}{\,#1}{#2}}
\newcommand{\massRatio}[1]{\sss{\zeta}{\,#1}}
\newcommand{\massRatioE}[2]{\ssss{\zeta}{\,#1}{#2}}
\newcommand{\powerScaleExponent}{k}

\begin{document}

\begin{frontmatter}

\title{Optimal mass and speed for interstellar flyby with directed-energy propulsion\tnoteref{t1}}
\tnotetext[t1]{Copyright\copyright 2020}

\author[1]{David G Messerschmitt}
\address[1]{University of Calfornia at Berkeley,
Department of Electrical Engineering and Computer Sciences, USA}

\author[2]{Philip Lubin}
\address[2]{University of California at Santa Barbara,
Department of Physics, USA}

\author[3]{Ian Morrison}
\address[3]{Curtin University, International Centre for Radio Astronomy Research, Australia}

\begin{abstract}

The design of mission scenarios for the flyby investigation of nearby star systems by probes launched using directed energy is addessed.
Multiple probes are launched with a fixed launch infrastructure, and download
of scientific data occurs following target encounter and data collection.
Assuming the primary goal is to reliably recover a larger 
volume of collected scientific data with a smaller data latency 
(elapsed time from launch to complete recovery of the data), 
it is shown that there is an efficient frontier where volume cannot be 
increased for a given latency and latency cannot be reduced for a given volume.
For each probe launch,
increasing the volume along this frontier is achieved by
increasing the probe mass, which results in a reduced probe speed.
Thus choosing the highest feasible probe speed generally does not achieve an efficient tradeoff
of volume and latency.
Along this frontier the total distance traveled to the completion of data download
does not vary significantly, implying that the download time duration is approximately
a fixed fraction of the launch-to-target transit time.
Due to longer propulsion duration
when probe mass is increased,
increasing data volume incurs a cost in the 
total launch energy expended, 
but with favorable economies of scale.
An important characteristic of any probe technology is the scaling law that
relates probe mass to transmit data rate,
as this affects details of the efficient frontier. 
\end{abstract}


\end{frontmatter}

\onecolumn

\begin{makefigurelist}

\section{Introduction}

The advancement of propulsion technology for interstellar spacecraft and probes generally emphasizes
achieving the maximum speed subject to cost constraints.
In other words, subject to budget constraints, the greater the speed the better.
This is a valid assumption in some cases, but not others.
Here we address the specific case of a probe flyby mission for scientific data collection
utilizing directed-energy propulsion,
and show that manipulation of the probe speed is a beneficial design freedom to employ
in addressing the needs of science investigators, who are the ultimate customers for the launch and
data-return technology.

The development here is specific to directed energy, which does not provide deceleration.
Thus it does \emph{not} apply to missions entering
an orbit or landing on a remote astronomical body,
nor to propulsion technologies other than directed-energy.
However, this special case illustrates how the ultimate purpose of a mission 
is an important consideration in the development of new propulsion technologies
as well as in mission design.
Parallel conclusions about matching spacecraft speed objectives to mission goals
have been encountered in interstellar deceleration and landing missions \cite{bond1986project,long2010project}.

Directed-energy propulsion is attractive for flyby because it eliminates the need for a probe to carry fuel
for propulsion, and this substantially reduces the mass at launch and thus enables
a post-launch speed that is a significant fraction of light speed $c$.
Such a flyby mission is assumed to consist of the phases illustrated in \figref{flybyMission}.
Since the directed-energy acceleration is short-lived relative to the
total mission duration, it is reasonable to assume that the probe
is ballistic; that is,
it travels at a constant speed throughout the mission.
The collection of a finite volume of scientific data (measured in bits) 
during target encounter is followed by downlink transmission
following encounter, so the probe continues its ballistic trajectory for the duration of 
communication downlink operation.

\incfig
	{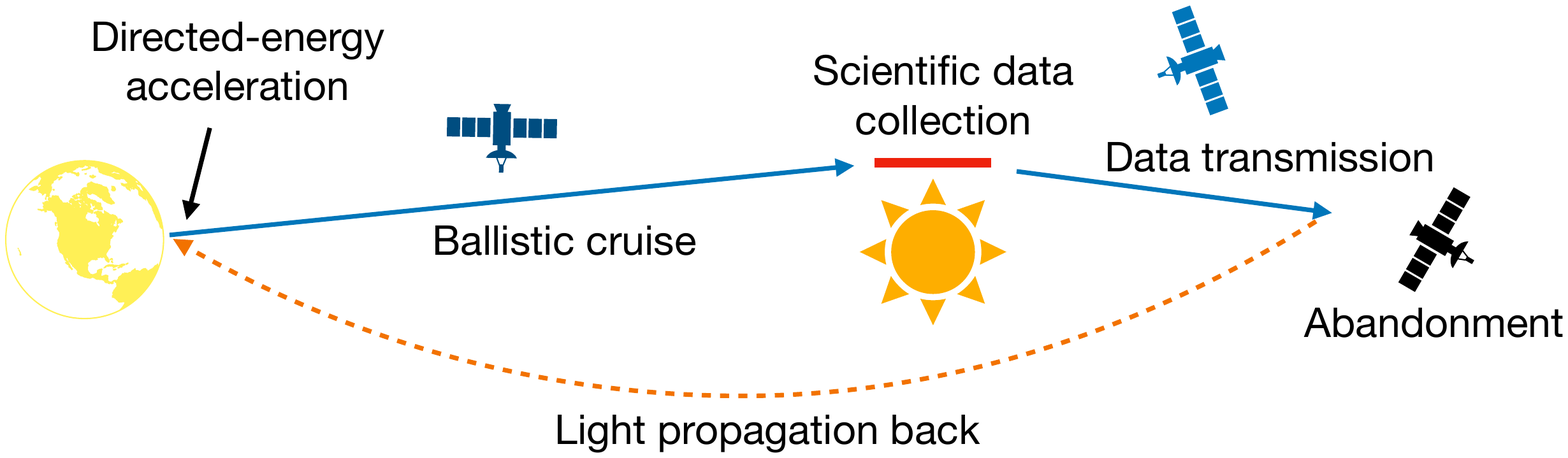}
	{
	trim=20 150 0 250,
    	clip,
    	width=.8\linewidth
	}
	{
	Illustration of the phases of a flyby mission for a probe propelled by
	directed energy from the launch site.
	The goal is to reliably recover
	the collected scientific data at the launch site.
	}

The major cost of a flyby mission is a directed-energy beamer at the launch site,
which may be terrestrial, on the moon, or a space platform.
Assuming that multiple probes are launched over time,
the launch energy is a significant cost.
Speeds in the range of $0.1 c$ to $0.2 c$
are often assumed\footnote{
 At these speeds relativistic effects are not very significant (on the order of a few percent),
 so for the purposes of
 analyzing a flyby mission we employ classical approximations throughout.
 }
because they permit travel times to the nearest
stars measured in decades, which is well within the duration of the typical career
of a space scientist or engineer.
An example of this is the ongoing StarShot project 
\cite{parkin2018breakthrough,RefNumber1015,RefNumber833,messer3}.
To achieve these speeds with credible cost goals requires a small probe mass (perhaps \ile{1-100\text{ g}}).
The major communication challenge is realizing a transmitter on the probe
with a small mass budget and which can communicate back from interstellar distances.

\section{Probe mass considerations in flyby missions}
\label{sec:massFlyby}

 Because the directed-energy launcher is usually assumed to be shared over multiple
launches, the launcher beam divergence and power are assumed to remain fixed,
and the probe speed can be manipulated by changing the probe mass
 \cite{RefNumber811,RefNumber1030,RefNumber833}.
 We show that performance metrics of direct interest to science investigators
 can be optimized by the choice of probe mass, since that mass indirectly affects
the instrumentation carried by the probe and the data rate available during downlink operation,
and because the probe speed affects the time available to perform science in the target vicinity
and the time available to downlink data for a given termination distance.

For a flyby mission,
the performance metrics of interest are listed in \tblref{metrics}
and the
design parameters available to manipulate those performance metrics are
listed in \tblref{parameters}.
 The primary purpose of a flyby mission is the collection of scientific data
in the vicinity of the target followed by the reliable recovery of that data
at or near the launch site.
The performance metrics of primary interest to scientist investigators
 are the data volume $\dataVolume$ and the  data latency $\dataLatency$.
In other words, ``how much data do we get back reliably, and how long do we have to wait for that data?''
 As will be seen, both these metrics are strongly influenced by the probe speed $\speedProbe{P}$.
 While domain scientists are usually not directly concerned with that speed,
 an exception is the impact on
 the time available for science investigations in the vicinity of the target.
 In this regard, slower (as advocated here) is always preferable.

 \doctable
	{metrics}
    {
    Flyby mission scientific performance metrics}
    {l p{10cm}}
    {
    \hline \\[-2ex]
   \textbf{Variable} & \textbf{Definition}
    \\ \hline \\ [-2ex]
    $\dataVolume$ & Total received volume of scientific data reliably recovered at Earth
     \\[2pt]
     $\dataLatency$ & Data latency = time elapsed from launch to reception of scientific data in its entirety
 \\[3pt] \hline
      }

 \doctable
	{parameters}
    {
    Flyby mission parameters}
    {l p{10cm}}
    {
    \hline \\[-2ex]
   \textbf{Variable} & \textbf{Definition}
    \\ \hline \\ [-2ex]
    $\timeCoord$ 
    & Classical coordinate time at launch site and at probe
    \\[2pt]
    $\timeTransmission$ & Time duration of transmission in coordinate time
\\[2pt]
$\mass{P}$ & Mass of probe, including sail, instrumentation, and communications
  \\[2pt]
    $\massRatio{P}$ & Mass ratio, equal to $\mass{P}/\mass{0}$, where $\mass{0}$ is a baseline value for mass
     \\[2pt]
$\speedProbe{P}$ & Ballistic probe coordinate speed, with value $\speedProbe{0}$ for \ile{\massRatio{P}= 1}
    \\[2pt]
    $\distanceTarget$ & Distance from launch to target star, and from probe transmitter to receiver at the start of downlink operation
    \\[2pt]
   $\rateData{\text{start}}$
   & Initial data rate at start of transmission, with value  $\rateData{0}$ for \ile{\massRatio{P} = 1}
     \\[2pt]
    $\powerScaleExponent$ 
    & Mass ratio to data-rate scaling exponent, so the data rate $\rateData{}$ scales by 
    \ile{\massRatioE{P}{\powerScaleExponent}}
 \\[3pt] \hline
      }

\subsection{Tradeoffs}

Two mission design parameters are
the  duration of downlink transmission $\timeTransmission$,
and the probe mass ratio $\massRatio{P}$, which is proportional to the probe mass
(where \ile{\massRatio{P}=1} for some baseline case).
If $\massRatio{P}$ is increased then it is appropriate to exploit that 
increased mass to increase the size of the
probe's sail, allowing the
duration of acceleration to increase accordingly (see \secref{kinematics}).
Despite this longer acceleration, the probe speed decreases from a baseline value
$\speedProbe{0}$ to \ile{\speedProbe{P} < \speedProbe{0}}.
The longer acceleration increases the energy expenditure, 
but that added energy expenditure is justified if the increased data volume
is substantial (see \secref{cost}).

Although the mission design parameters \ile{\{\massRatio{P},\timeTransmission\}} can be varied to 
manipulate the mission performance metrics \ile{\{\dataVolume,\dataLatency\}},
they should not be chosen arbitrarily.
Rather, they should be
jointly optimized to achieve the most favorable \ile{\{\dataVolume,\dataLatency\}}
tradeoff.
The impact of \ile{\{\massRatio{P},\timeTransmission\}} on \ile{\{\dataVolume,\dataLatency\}} is slightly complicated,
but can be summarized as:
\begin{itemize}
\item
A larger $\massRatio{P}$ results in a
smaller $\speedProbe{P}$.
\item
 This increases the travel time to the target, and this increases
 $\dataLatency$.
\item
This results in a smaller accumulation of distance-squared propagation delay,
and thus allows $\rateData{}$ to decrease more slowly during downlink transmission,
which in turn increases $\dataVolume$ (see \secref{volLatency}).
\item
For fixed $\timeTransmission$, the probe has traveled less distance from the target
during downlink operation, the maximum propagation delay back to the launch site
is smaller, and this reduces $\dataLatency$.
\item
A larger probe mass budget for communications (including its electrical power generation)
allows $\rateData{}$ to be increased (see \secref{volLatency}), and this increases $\dataVolume$.
\end{itemize}
While the travel-time increase is deleterious, all the other impacts are beneficial.
Trading these off leads to an optimum point.
We now summarize the conclusions of this optimization, 
followed by a supporting analysis in \secref{massRatioAnalysis}.

\section{Optimal volume-latency tradeoff}
\label{sec:volumeLatencyTradeoff}

The design of a data link, which conveys data reliably from probe to launch site,
involves a number of interacting considerations such as wavelength,
transmit aperture, receive collector, background radiation,
modulation, and coding.
For the purposes of mission design,
all these considerations can be wrapped into a single parameter: 
a baseline data rate $\rateData{0}$ at the beginning of
downlink operation assuming \ile{\massRatio{P}=1}.
The total data volume \ile{\dataVolume \propto \rateData{0}}, and thus
it is convenient to use the normalized volume \ile{\dataVolumeNorm}
(which is dimensioned in time) as a performance metric to guide the
choice of \ile{\{\massRatio{P},\timeTransmission\}}.

\subsection{Efficient frontier}
\label{sec:efficientFrontier}
 
 The tradeoff between $\dataVolumeNorm$ and $\dataLatency$
 (first explored in \cite{RefNumber833}) is plotted in \figref{efficientFrontierVsFixedMassRatio}.
 There exists a feasible region of operation \ile{\{\dataVolumeNorm,\dataLatency\}}, which
 is shaded in \figref{efficientFrontierVsFixedMassRatio}.
 Points on the lower boundary of
 this region, called the \emph{efficient frontier}\footnote{
 This terminology is borrowed from a similar concept in financial portfolio theory \cite{merton1972analytic}.
 It is a special case of the \emph{Pareto frontier} (Pareto optimization is widely employed in
 various engineering disciplines \cite{jakob2014pareto}).
 },
 constitute all the advantageous mission operating points.
That boundary yields the smallest possible $\dataLatency$ for a given $\dataVolumeNorm$,
or the largest possible $\dataVolumeNorm$ for a given $\dataLatency$.

Choice of a mission operation point somewhere on the efficient frontier provides flexibility
in setting mission priorities.
There are several compelling reasons to consciously select different
operating points along the efficient frontier for different missions 
sharing a common launch infrastructure:
\begin{itemize}
\item
 Mission designers can consciously
 prioritize large $\dataVolumeNorm$ or small $\dataLatency$.
 \item
 Different probes may carry different types of instrumentation,
 and these impose different mass and data volume requirements.
 \item
 There will likely be an evolution of probe technology over time.
 Early probes may emphasize technology validation
 with low $\dataLatency$ (and hence small $\dataVolumeNorm$),
 while later probes may emphasize scientific return with larger $\dataVolumeNorm$
 (and hence larger  $\dataLatency$).
 \item
 There may be missions to different targets at different distances
(within the solar system and interstellar), significantly changing the
 possible range of $\dataLatency$.
 \end{itemize}
 Generally mission designers will seek to maximize an objective function that
 combines volume and latency objectives.
 Only points along the efficient frontier need be considered in any such optimization.
 
 Also illustrated in \figref{efficientFrontierVsFixedMassRatio} as the dashed curve is
 a set of possible mission operation points when
a baseline value \ile{\massRatio{P}=1}  is chosen and only $\timeTransmission$ is varied.
 This arbitrary choice of  \ile{\massRatio{P}} permits operation at exactly 
 one point on the efficient frontier through a
 judicious choice of $\timeTransmission$.
More generally, achieving an arbitrary operating point on the efficient frontier requires a
coordinated choice of \ile{\{\massRatio{P},\timeTransmission\}} rather than constraining
$\massRatio{P}$ in this manner.

\incfig
	{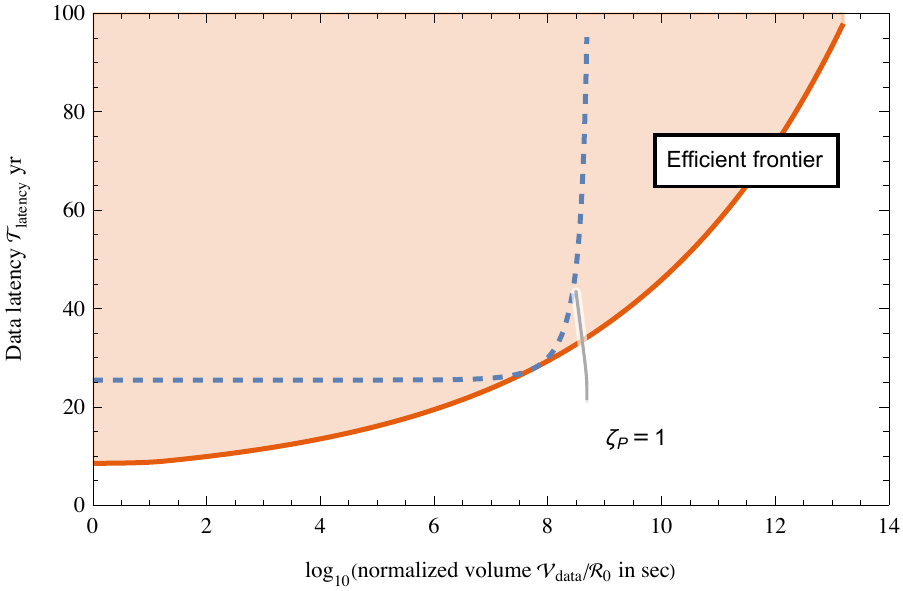}
	{
	trim=0 0 0 0,
    	clip,
    	width=0.7\linewidth
	}
	{
Plots of data latency $\dataLatency$ (in years) vs the log of the normalized data volume
	$\dataVolumeNorm$ (in seconds) where 
	$\rateData{0}$ is the data rate (in bits per second) 
	at the beginning of downlink transmission
	(data rate declines from there as the square of propagation distance) for mass ratio \ile{\massRatio{P}=1}.
	$\dataVolume$ (in bits) is found by multiplying by the assumed value for $\rateData{0}$.
	Any volume-latency mission operating point within the shaded region is feasible.
	The lower boundary of this region, called the efficient frontier, is an efficient operating point in the 
	sense of maximizing the volume for a given latency,
	or minimizing latency for a given volume.
	The set of operation points obtained by fixing \ile{\massRatio{P} = 1}
	and varying downlink operation duration $\timeTransmission$ are shown as a dashed curve.
}

\subsection{Origin of efficient frontier}

Additional insight into the efficient frontier follows from examining the fixed $\massRatio{P}$
dashed curve.
Its general shape follows from two asymptotes as illustrated in \figref{fixedMassPlots}a.
For small $\dataVolumeNorm$ the horizontal asymptote is due to
 the minimum possible data latency $\dataLatency$ when \ile{\timeTransmission \to 0}.
In this event downlink operation duration is not a factor and
$\dataLatency$  becomes dominated by the sum of launch-to-target transit time and signal propagation
time back to the receiver.
Similarly, $\dataVolumeNorm$ is bounded from above by a 
vertical asymptote, which follows from the maximum $\dataVolume$ as \ile{\timeTransmission \to \infty}.
The increasing distance
of the probe during downlink operation reduces the data rate $\rateData{}$ as
distance-squared, and the integral of $\rateData{}$ is finite even as $\timeTransmission$
becomes arbitrarily large.
The efficient frontier is achieved by a judicious choice of an appropriate $\timeTransmission$ 
 intermediate to these two asymptotes.
Varying the fixed value of $\massRatio{P}$
results in a family of curves as illustrated in \figref{fixedMassPlots}b.
The efficient frontier is the lower envelope of this family of curves.

\incfig
	{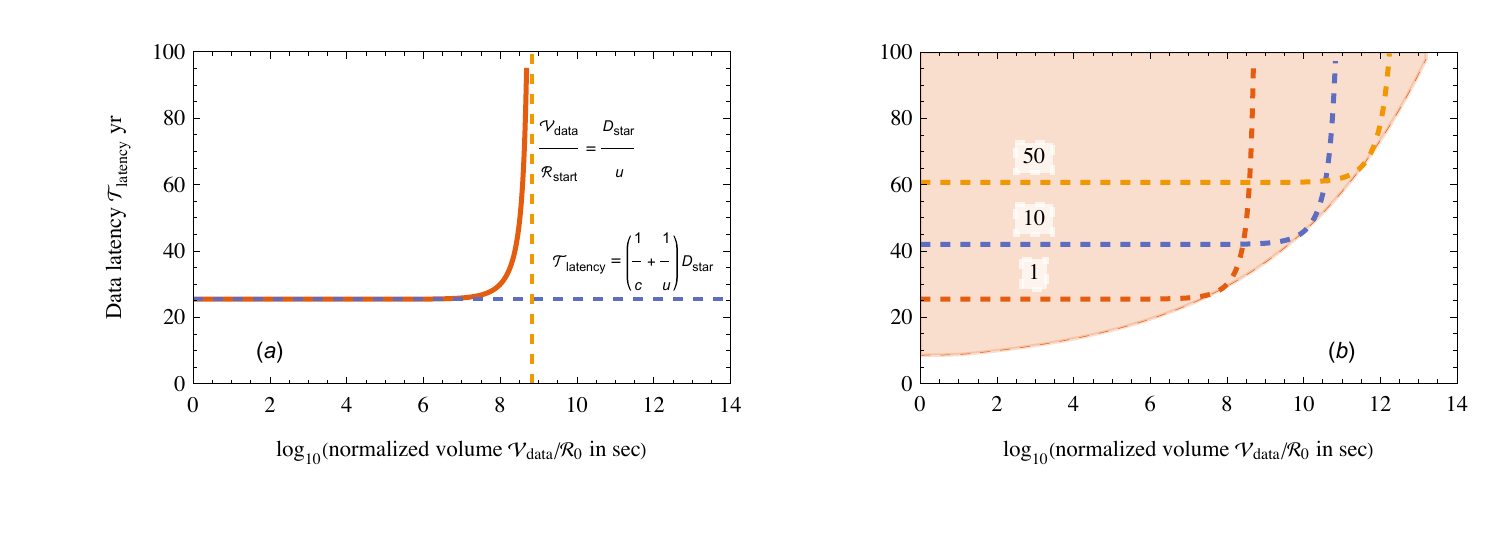}
	{
	trim=20 20 0 0,
    	clip,
    	width=1\linewidth
	}
	{
An illustration of the fixed mass ratio $\massRatio{P}$ operating points as the
	downlink operation duration $\timeTransmission$ is varied.
	(a) The dashed curve in \figref{efficientFrontierVsFixedMassRatio} is interpreted in terms
	of its two asymptotes, one determined by the minimum possible $\dataLatency$ and the
	other determined by the maximum possible $\dataVolumeNorm$.
(b) A repeat of \figref{efficientFrontierVsFixedMassRatio} showing \ile{\massRatio{P} \in \{1,10,50\}}.
	Increasing $\massRatio{P}$ results in an increase in the minimum latency asymptote
	(due to a reduction in the probe speed)
	as well as an increase in the $\dataVolumeNorm$ asymptote
	(due to a slower reduction in $\rateData{}$).
}

\subsection{Data rate scaling law}

The details of the efficient frontier are affected by the relationship between
$\massRatio{P}$ and $\rateData{}$.
Actually $\rateData{}$ is directly related to a second mass ratio $\massRatio{C}$,
which is the factor by which the mass of the communications subsystem is increased
even as the mass of the entire probe is increased by $\massRatio{P}$.
It is shown in \secref{allocation} that under two distinct but reasonable sets of
assumptions \ile{\massRatio{C} \ge \massRatio{P}}, so the mass-ratio budget available
for communications is at least as generous as for the probe in its entirety.
When a communications subsystem is much lighter than the remainder of the
payload, and a disproportionate part of any mass increase is devoted to communications,
 $\massRatio{C}$ can potentially be much larger than $\massRatio{P}$.

$\rateData{}$ is proportional to the product of transmit power and the transmit aperture area,
among other factors (such as receive collector area).
In view of this,
there are two distinct ways in which \ile{\massRatio{C} > 1} can be exploited
to increase $\rateData{}$:
\begin{description}
\item{\textbf{Increased electrical power:}}
Electrical power can be increased by $\massRatio{C}$.
As an existence proof, for any given electrical generator technology the replication of $J$ such generators
results in a mass and power that is a factor of $J$ larger.
Consolidating these generators into fewer and larger is worthwhile only if the outcome is
a materially improved power-to-mass ratio.
\item{\textbf{Increased transmit aperture area:}}
If the radiation area of a transmit aperture is increased by a factor of $J$, then its mass
may need to be no more than $J$ times larger.
However, the available aperture fabrication technology may not offer quite this favorable a tradeoff,
if for example additional mass may be necessary as a means to strengthen the larger structure.
There also may be other limitations on transmit aperture area, such as the available pointing accuracy.
The transmit aperture area should be matched to that pointing accuracy, so that the
beam divergence is sufficiently large (aperture area sufficiently small) to cover the receiver
in the presence of the worst-case pointing offset.
\end{description}
These observations suggest that the data rate $\rateData{\text{start}}$ at the start of downlink operation
can be related to $\massRatio{C}$ through
\begin{equation}
\label{eq:scaleExp}
\rateData{\text{start}} = \massRatioE{C}{\powerScaleExponent} \,\rateData{0}
\end{equation}
and $1 \le \powerScaleExponent \le 2$ is a transmit power \emph{scaling exponent}.
The  \ile{\powerScaleExponent = 1} case would apply when the transmit aperture
size is not increased at all, and \ile{\powerScaleExponent = 2} would apply if the benefits of increased mass
on both power and area are fully exploited.
The efficient frontier for these extreme cases is compared in \figref{efficientFrontierVsScaling}.
Not surprisingly \ile{\powerScaleExponent = 2} is prefered because it can achieve one to two orders of
magnitude larger $\dataVolumeNorm$.

\incfig
	{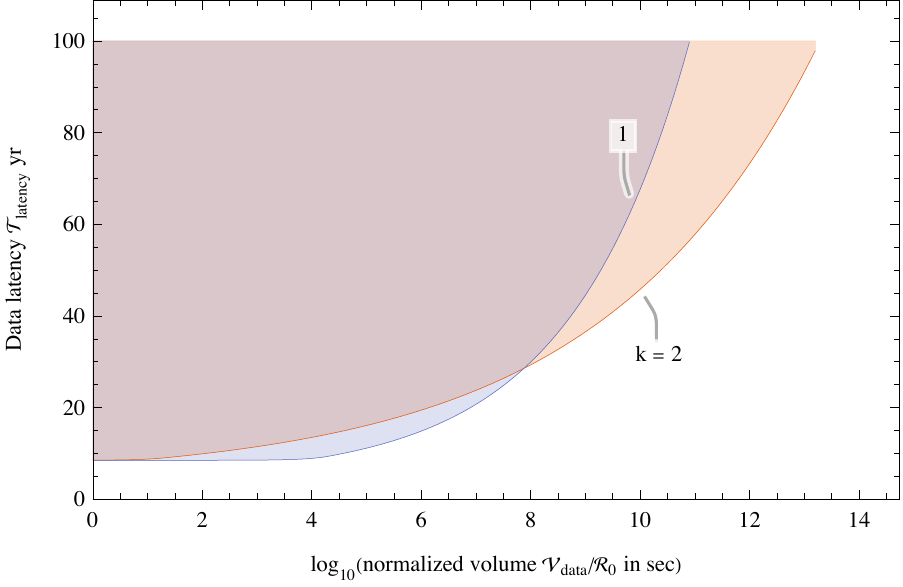}
	{
	trim=0 0 0 0,
    	clip,
    	width=0.7\linewidth
	}
	{
	The efficient frontier in \figref{efficientFrontierVsFixedMassRatio} is compared for 
	scaling exponents \ile{\powerScaleExponent = 2} and \ile{\powerScaleExponent = 1}.
	The proportional case (\ile{\massRatio{C} = \massRatio{P}}) is assumed.
	}

\subsection{Other dependent mission parameters}

The mass ratio $\massRatio{P}$ (in conjunction with $\powerScaleExponent$) is the 
independent mission parameter that affects the
operating point along the efficient frontier.
Since a specific point on the efficient frontier corresponds to a specific choice of
\ile{\{\massRatio{P},\timeTransmission\}}, other mission parameters that must be chosen are dependent on this.

 The value of $\massRatio{P}$ is shown in \figref{massSpeedVsScaling}a for the assumptions
 underlying \figref{efficientFrontierVsScaling}.
 This has the secondary effect of determining the probe speed $\speedProbe{P}$
 as shown in \figref{massSpeedVsScaling}b.
 With increasing $\dataVolumeNorm$, the mass ratio increases and the speed decreases.
 Increasing $\dataVolumeNorm$ and operating on the efficient frontier
 also requires an increase in transmission time $\timeTransmission$, 
 as shown in \figref{timeVsScaling}a.
 However, the transit time from launch to target also increases due to lower probe speed,
 and the ratio of $\timeTransmission$ to that transit time remains relatively constant
 as shown in \figref{timeVsScaling}b.
 Since both transit time and downlink operation duration scale with probe speed,
 the actual distance traveled by the probe during downlink operation doesn't
 vary much at different points on the efficient frontier.

\incfig
	{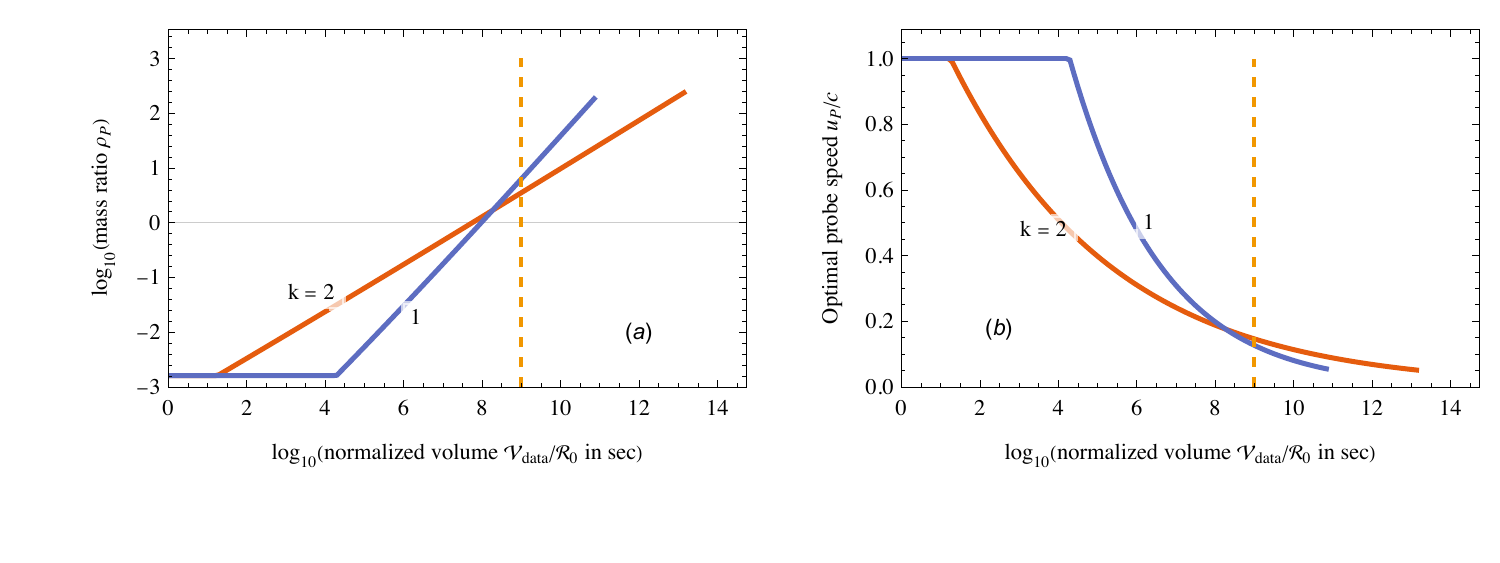}
	{
	trim=20 20 0 0,
    	clip,
    	width=1\linewidth
	}
	{
Assuming operation on the efficient frontier in \figref{efficientFrontierVsScaling},
	a log plot of the optimal mass ratio $\massRatio{P}$ in (a) and
	a plot of the resulting probe speed as a fraction of the light speed in (b)
	assuming baseline speed \ile{\speedProbe{0}=0.2 c} at \ile{\massRatio{P}=1}.
	The two scaling exponents \ile{\powerScaleExponent \in \{1,2\}} are plotted and labeled.
	The vertical dashed line illustrates one data volume of interest,
	which is \ile{\dataVolume=10^9\text{ bits}} (\ile{1\text{ Gb}}) at 
	a baseline data rate \ile{\rateData{0}=1\text{ bits s}^{-1}} at the beginning of downlink operation.
	\ile{\massRatio{P} \ll 1} achieves low data latency by increasing the probe speed and thus reducing the transit
	time to the target (at the expense of a low data volume).
	\ile{\massRatio{P} \gg 1} achieves a larger data volume by 
	reducing the probe speed (resulting in a larger data latency).
}

\incfig
	{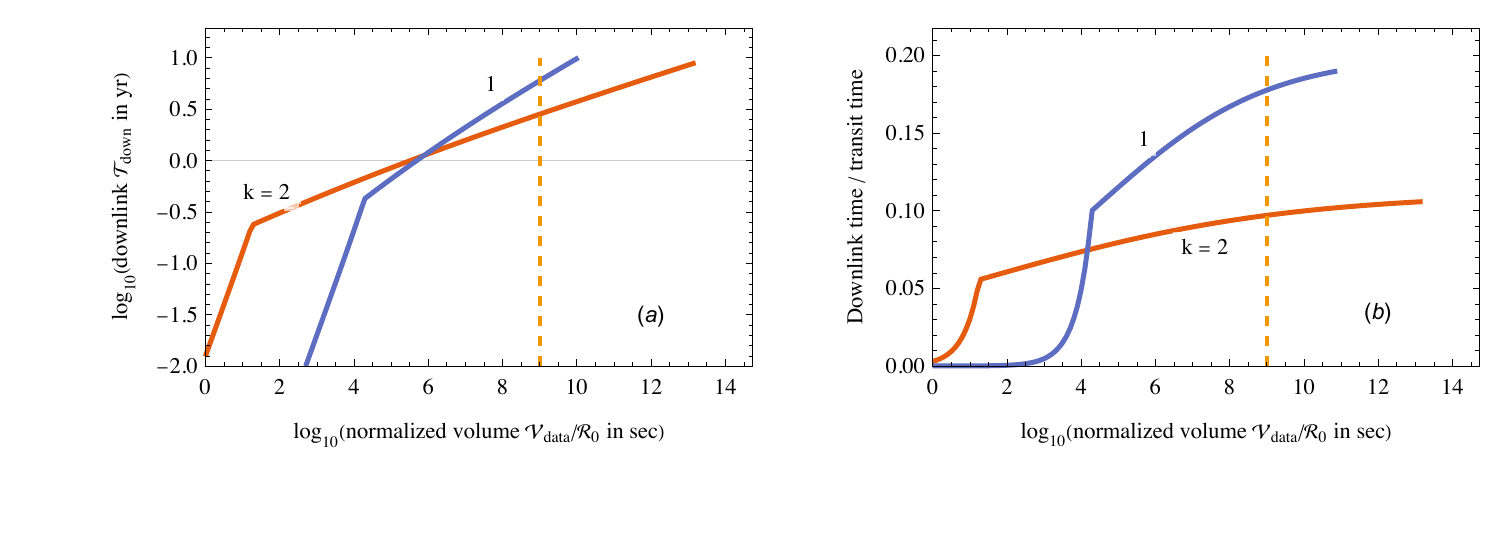}
	{
	trim=20 20 0 0,
    	clip,
    	width=1\linewidth
	}
	{
Under the same conditions as \figref{massSpeedVsScaling},
	a plot of (a) the log of
	the optimal downlink operation duration $\timeTransmission$ and (b)
	 the ratio of $\timeTransmission$ to the transit time from launch to target.
	A rule of thumb is that for \ile{\massRatio{P} \gg 1},this ratio should be about 9\% for
	\ile{\powerScaleExponent =2} and 17\% for \ile{\powerScaleExponent =1}.
	The distance flown during downlink operation has the same relationship to the
	launch-to-target distance.
}

The overall conclusion is that the squared-distance dependency of $\rateData{}$ is the
dominant consideration.
That results in a distance traveled from launch to the end of downlink transmission that is
relatively invariant across different operating points on the efficient frontier.
The primary tool for adjusting the volume-latency tradeoff on the frontier is 
probe speed rather than distance traveled during downlink transmission.

\section{Analysis}
\label{sec:massRatioAnalysis}

An analytic treatment of the volume-latency tradeoff provides additional insight.

\subsection{Directed energy kinematics}
\label{sec:kinematics}

The total probe mass $\mass{P}$ can be broken into constituents as
\begin{equation}
\label{eq:massCategories}
\mass{P} = \mass{C} + \mass{E} + \mass{S}
\,,
\end{equation}
where $\mass{C}$ is the communications subsystem, $\mass{S}$ is the sail,
and $\mass{E}$ is everything else (including attitude control, pointing, and scientific instrumentation).
Electric power generation serves communication, scientific instrumentation and attitude control,
but communication and instrumentation can share generation capacity
since they do not need to operate concurrently.

The kinematics of a directed-energy launch was studied in \cite{RefNumber811} 
and is reviewed in \secref{beamerSailAnalysis}.
This establishes that the
$\mass{S}$ should always make up exactly half of $\mass{P}$
in order to achieve the maximum probe speed during the ballistic phase of the mission.
With this optimum $\mass{S}$, the probe speed scales as
\ile{\speedProbe{P} \propto {\massRatio{P}}^{-1/4}}.
The total launch energy scales as
\ile{\energyLaunch \propto {\massRatio{P}}^{3/4}}.
Thus the launch energy increases as the probe speed decreases because the
directed-energy beam takes longer to reach the diffraction limit with a larger sail.
\begin{example}
When \ile{\massRatio{P} = 16}, the probe speed $\speedProbe{P}$ is reduced by a factor of 
\ile{16^{1/4} = 2}
and the launch energy $\energyLaunch$ is increased by a factor of \ile{16^{3/4} = 8}.
For scaling exponent \ile{k=2} the data rate $\rateData{\text{start}}$ immediately
following encounter is increased by a factor of \ile{16^2 = 256},
which increases the normalized data volume $\dataVolumeNorm$ by that same factor.
\end{example}

\subsection{Mass allocation}
\label{sec:allocation}

Communication mass ratio $\massRatio{C}$ may beneficially be larger than $\massRatio{P}$,
as now discussed.
Assume the baseline probe masses
\ile{\{ \mass{P,0}, \mass{C,0}, \mass{E,0}, \mass{S,0}\}}
associated with the mass categories in \eqnref{massCategories},
and the variation of mass across different probe missions can be expressed in the mass ratios
\begin{equation}
\massRatio{P} = \frac{\mass{P}}{\mass{P,0}} \,,\text{  and }
\massRatio{C} = \frac{\mass{C}}{\mass{C,0}}
\,.
\end{equation}
The kinetic law  \ile{\mass{P} = 2 \mass{S}} is also assumed,
with the result that $\massRatio{P}$ determines the probe speed $\speedProbe{0}$
as described in \secref{kinematics}.
The other mass ratio $\massRatio{C}$ determines how much resource can be devoted to
achieving an initial data rate $\rateData{\text{start}}$.
Thus the relationship between $\massRatio{P}$ and $\massRatio{C}$ is significant.
We address this question under two alternative assumptions:
\begin{description}
\item{\textbf{Proportional masses:}}
As $\mass{P}$ is varied, \ile{\mass{C} \propto \mass{E}}, in which case \ile{\massRatio{C}=\massRatio{P}}.
Thus an increase in $\mass{P}$ provides an equivalent benefit (in terms of mass ratio) to the 
communications and instrumentation subsystems.
Not only can $\rateData{\text{start}}$ be increased, but also the mass devoted to instrumentation can be increased.
These two increases may go hand in hand, if for example a more massive instrumentation benefits from a larger
data volume.
\item{\textbf{Fixed mass:}}
As $\mass{P}$ is varied $\mass{E}$ is kept fixed, so the entirety of a probe mass $\mass{P}$ increase
is devoted to increasing $\mass{C}$.
In this case the fraction of the mass devoted to communications becomes relevant.
An additional mass ratio is defined,
\begin{equation}
\gamma = \frac
	{\mass{C,0}}
	{\mass{C,0} + \mass{E}}
	\,.
\end{equation}
\end{description}
Overall we find that
\begin{equation}
\label{eq:massRatioC}
\massRatio{C} = \begin{cases}
 \massRatio{P}
 \,, & 
 \text{proportional}
 \\
1 + \frac{\massRatio{P} - 1}{\gamma}
\,,
&
\text{fixed}
\end{cases}
\,.
\end{equation}
Since \ile{0 < \gamma < 1}
it follows that \ile{\massRatio{P} \le \massRatio{C}},
with equality at the limit as \ile{\gamma \to 1}.
For low-mass probes we would expect a $\gamma$ to be relatively large,
say \ile{\gamma \approx 0.5}, which results in smaller impact on $\massRatio{C}$
than for a massive spacecraft.
The fixed case is always more favorable to
communications since the entirety of any mass increase benefits the communications subsystem.
To the extent that $\massRatio{C}$ is larger than $\massRatio{P}$,
this benefits $\dataVolumeNorm$ as illustrated in
\figref{efficientFrontierVsCommMass}.

\incfig
	{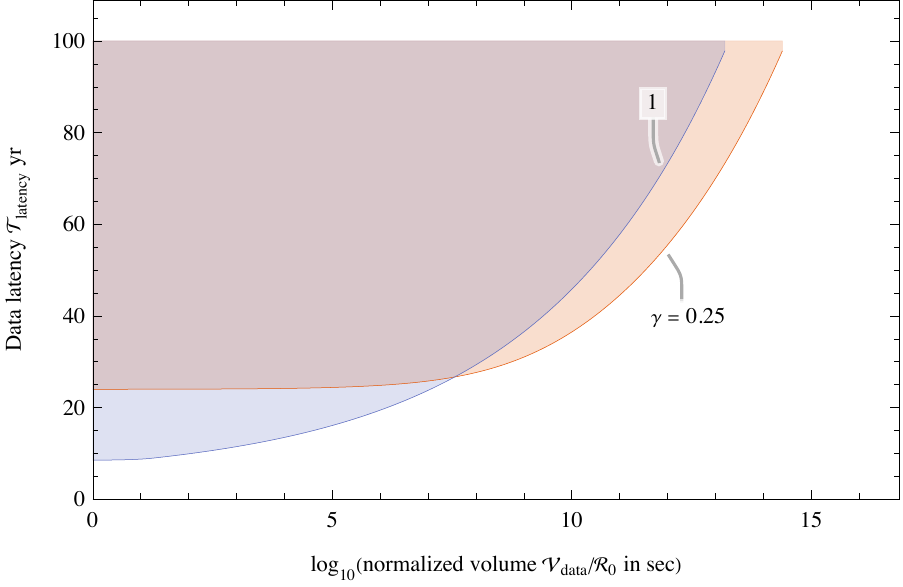}
	{
	trim=0 0 0 0,
    	clip,
    	width=0.7\linewidth
	}
	{
	Repeating \figref{efficientFrontierVsScaling},
	and the scaling exponent \ile{\powerScaleExponent =2},
	the efficient frontier is plotted for two values \ile{\gamma \in \{1,0.25\}}.
	The \ile{\gamma=1} case assumes that the mass change devoted to communications
	is proportional to the total mass change.
	The \ile{\gamma=0.25} case assumes that the 
	total non-communication mass (exclusive of sail) is $3\times$ larger than the communications
	subsystem mass at the baseline,
	and further that non-communication mass remains fixed as any 
	mass change along the efficient frontier is allocated entirely to communications. 
	This \ile{\gamma=0.25} case magnifies the impact of that mass change on $\dataVolumeNorm$
	because of the disproportionate effect on $\rateData{\text{start}}$.
	}

\subsection{Determination of efficient frontier}

The efficient frontier is found by numerical minimization of $\dataLatency$ with respect to $\massRatio{P}$
for each value of $\dataVolumeNorm$ of interest (see \secref{volLatency}).
An approximation that avoids the numerical minimization follows by assuming
(based on the numerical results of \figref{timeVsScaling}b)
that the downlink operaton time $\timeTransmission$ is
a constant fraction of the launch-to-encounter time
(see \secref{appxEfficientFrontier}).

\subsection{Launch energy cost and economies of scale}
\label{sec:cost}

Although the launch infrastructure remains fixed, the variable launch costs increase
as $\dataVolumeNorm$ increases along the efficient frontier.
This is because
an increase in $\massRatio{P}$ implies an increase in launch energy in spite of the lower probe speed
(see \secref{beamerSailAnalysis}).
Launch energy $\energyLaunch$ is plotted in \figref{launchEnergy}a.

Also shown in \figref{launchEnergy}b is the ratio of $\energyLaunch$ to $\dataVolumeNorm$,
which decreases steadily.
If we view $\energyLaunch$ as a primary variable cost of scientific data return and a larger $\dataVolumeNorm$
as the reward for that expenditure, then data return exhibits significant economies of scale.
In particular, increasing $\massRatio{P}$ in the interest of a larger $\dataVolume$ incurs a lower cost
than launching duplicative less massive probes to achieve the same overall $\dataVolume$.
Of course other system objectives such as reliability and diversity of scientific instrumentation
should be taken into account as well.

\incfig
	{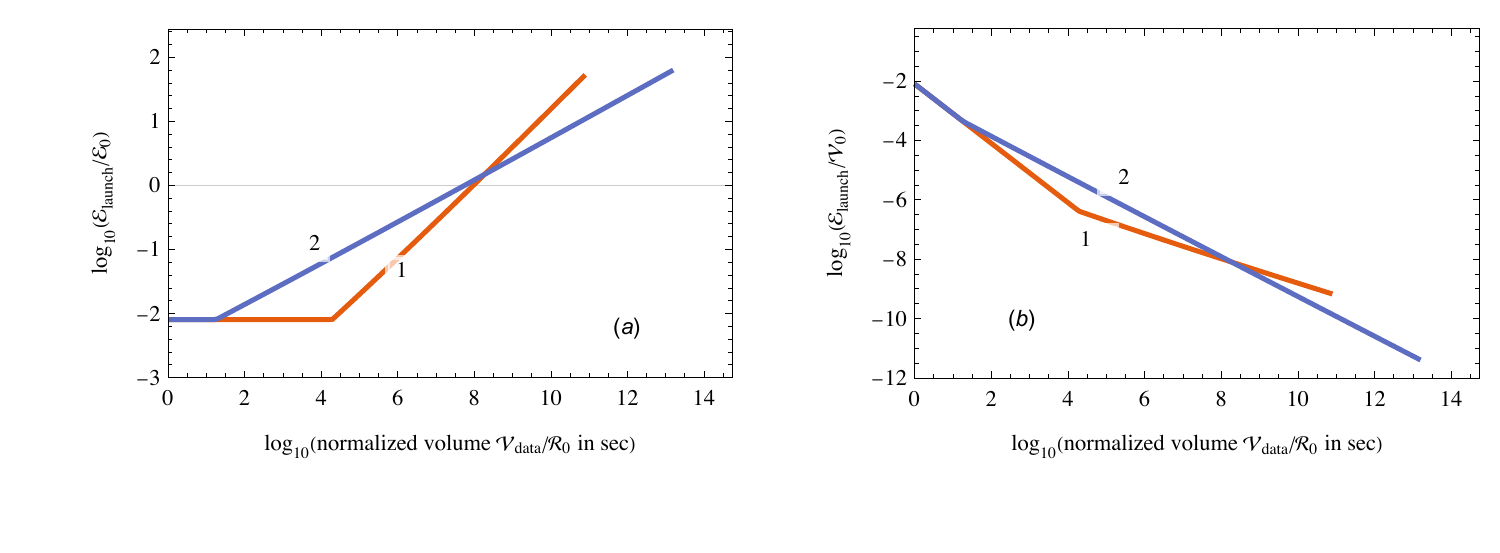}
	{
	trim=20 20 0 0,
    	clip,
    	width=1\linewidth
	}
	{
For operation on the efficient frontier in \figref{efficientFrontierVsScaling},
	a log plot of the relationship of launch energy $\energyLaunch$ to normalized data volume
	$\dataVolumeNorm$.
	The two scaling exponents \ile{\powerScaleExponent \in \{1,2\}} are plotted and labeled.
	(a) $\energyLaunch$ increases with data volume because of the larger mass probe
	that is necessitated to achieve that volume.
	(b) The ratio of $\energyLaunch$ to $\dataVolumeNorm$ decreases steadily, demonstrating
	that additional $\dataVolumeNorm$ comes at a lower and lower incremental cost in terms of
	launch energy.
}

\section{Conclusions}

This study undermines
any presumption for interstellar missions that the maximum probe speed should be achieved.
To ascertain the best choice of speed,
the needs of the ultimate stakeholders
should be assessed, leading to optimized mission parameters.
For the mission scenario considered here, with an emphasis on performance
parameters ultimately of interest to domain science investigators in a directed-energy flyby mission,
the conclusion is that any concrete choice of probe speed can achieve
only a single point on the efficient frontier, and achieving other
optimal volume-latency tradeoffs requires an appropriate choice of probe mass ratio and thus speed.
Further, this optimization determines other mission parameters such
as transmission time, and secondarily the launch energy requirement.
The remaining degree of freedom is the data volume-latency tradeoff, which is achieved
by moving along the efficient frontier.
An additional consideration is the instrumentation (as to both  mass and electrical
power requirements)
which may also influence the choice of probe mass as well.

While the efficient frontier is a universal concept,
its particulars are dependent on the mass-to-data-rate scaling law, which is an important characteristic
of any assumed probe technology and design, and is
also affected by the choice of instrumentation.
Also, the results and conclusions apply to directed-energy propulsion with
a fixed launcher infrastructure, 
with the only launch parameter dependent on
probe mass being the duration of launch acceleration.
This is a natural assumption for a launcher that is shared among multiple
probes with heterogenous instrumentation, data volume, and latency preferences.

\section{Acknowledgements}

The contribution of anonymous reviewers to improvement in this paper is appreciated.

PML funding for this program comes from NASA grants NIAC Phase I DEEP-IN ? 2015 NNX15AL91G and NASA NIAC Phase II DEIS 2016 NNX16AL32G and the NASA California Space Grant NASA NNX10AT93H and the Emmett and Gladys W.  Technology Fund as well as from Limitless Space Institute and Breakthrough Initiatives.

\appendix

\section{Launcher-sail analysis}
\label{sec:beamerSailAnalysis}

We employ a classical model, which gives results approximating a relativistic model
\cite{RefNumber851} as long as the probe speed remains in the sub-relativistic regime
(\ile{\speedProbe{P} < 0.5 c} or so).
At higher speeds a relativistic model is indicated.

Assume that the beamer directs a fixed power at the sail for a time period
$\timeLaunch$, and all that power
is reflected by the sail (that is the sail remains within the diffraction limit throughout the acceleration).
Then the force $F$ and acceleration $a$ on the sail both remain constant throughout acceleration.
Assume the total mass of the probe is $\mass{P}$, and this includes the sail mass $\mass{S}$.
The kinematics can then be summarized by
\begin{equation*}
F = \mass{P} \,a \,,\ \ 
\big( a \timeLaunchSquared / 2 \big) \propto \sqrt{\mass{S}} \,,\ \ 
\speedProbe{P} = a\, \timeLaunch \,.
\end{equation*}
The distance over which acceleration occurs
(until the diffraction limit is reached) is
proportional to the diameter of the sail, which in turn is proportional to $\sqrt{\mass{S}}$,
leading to the second equation.
Solving for $\speedProbe{P}$ and differentiating establishes that the maximum ballistic speed
$\speedProbe{P}$ is achieved for the choice \ile{2 \mass{S}= \mass{P}}.
Adopting this value for $\mass{S}$, the result is that \ile{\speedProbe{P} \propto {\mass{P}}^{-1/4}}.
A further conclusion is that \ile{\timeLaunch \propto \massE{P}{3/4}}, 
and thus the launch energy \ile{\energyLaunch \propto \massRatioE{P}{3/4}}.

\section{Volume-latency relations}
\label{sec:volLatency}

The achievable $\rateData{}$ for any efficient communication link design is proportional to received power,
which follows a distance-squared law.
Thus for a given starting data rate $\rateData{\text{start}}$, the 
best achievable data rate as a function of coordinate time
$\timeCoord$ decreases as,
\begin{equation*}
\label{eq:powerVsDistance}
\frac{\rateData{} \big[ \timeCoord \big]}{\rateData{\text{start}}} =  
\left( \frac{\distanceTarget}{\distanceTarget+\speedProbe{P} \timeCoord} \right)^2 
\,.
\end{equation*}
The total data volume follows by integration,
\begin{equation}
\label{eq:volume}
\frac{\dataVolume}{\rateData{\text{start}}}  = 
\int_{\sss{t}{1} = 0}^{\timeTransmission}
\frac{\rateData{} \big[ \timeCoord \big]}{\rateData{\text{start}}} \cdot \text d \timeCoord = 
\frac{\distanceTarget \timeTransmission}{\distanceTarget+\speedProbe{P}\, \timeTransmission}
\ \xrightarrow[\timeTransmission \to \infty]{} \ \frac{\distanceTarget}{\speedProbe{P}} \,.
\end{equation}
There are two profound implications.
First, $\dataVolume$ in \eqnref{volume} increases as $\speedProbe{P}$ decreases because of the 
slower rate of increase in propagation loss.
Second,
$\dataVolume$ is bounded  even when the energy available for transmission is unlimited.

The data latency is given by
\begin{equation}
\label{eq:latency}
\dataLatency = 
\big( \distanceTarget + \speedProbe{P} \timeTransmission \big) 
\left(\frac{1}{\speedProbe{P}} + \frac{1}{\speedLight}\right)
\,.
\end{equation}
where the first term is the total distance flown to the end of downlink operation and
the second term takes into account both the transit time to that distance and the signal
propagation time back from that distance.

\eqnref{volume} and \eqnref{latency} can then be solved simultaneously to obtain 
\ile{\{\dataLatency, \timeTransmission\}} 
as a function of $\dataVolumeNorm$,
\begin{equation}
\label{eq:solution}
\dataLatency = \frac
{\distanceTargetSquared \rateData{\text{start}}}
{\distanceTarget \rateData{\text{start}}-\speedProbe{P} \dataVolume}
\left( \frac{1}{\speedProbe{P}} + \frac{1}{\speedLight} \right)
,\ 
\timeTransmission = \frac
{\distanceTarget \dataVolume}
{\distanceTarget \rateData{\text{start}} - \speedProbe{P} \dataVolume}
\,.
\end{equation}
The domain of applicability was determined in \eqnref{volume}.
The scaling laws in \eqnref{scaleExp}, \eqnref{massRatioC}, and from \secref{kinematics} 
can be substituted.
The efficient frontier is then obtained by numerically minimizing $\dataLatency$ with respect to $\massRatio{P}$
for each value of $\dataVolumeNorm$ of interest.
The compatible value for $\timeTransmission$ then follows directly from \eqnref{solution}.

\section{Approximation to efficient frontier}
\label{sec:appxEfficientFrontier}

Examining \figref{timeVsScaling}b, the numerical minimization in determining the efficient frontier
can be avoided by choosing an approximate downlink operation time
\begin{equation}
\label{eq:constTransTime}
\timeTransmission \simeq b \cdot \distanceTarget/\speedProbe{P}
\,,
\end{equation}
where \ile{b\approx 0.09} for \ile{\powerScaleExponent=2} 
and  \ile{b\approx 0.17} for \ile{\powerScaleExponent=1}.
The resulting efficient frontier 
(as a curve parameterized by $\massRatio{P}$) is
\begin{equation}
\dataVolumeNorm \approx
\frac{b\, \distanceTarget {\massRatio{P}}^{1/4} \left(\frac{\gamma
   +\massRatio{P} -1}{\gamma }\right)^k}{(b+1)\, \speedProbe{0}}
 ,\ 
\dataLatency \approx
(b+1)\, \distanceTarget
   \left(\frac{1}{c}+\frac{{\massRatio{P}}^{1/4}}{\speedProbe{0}}\right)
 \,.
\end{equation}
The accuracy of
this approximation can be verified by numerical comparison.

\end{makefigurelist}


\begin{thebibliography}{10}
\expandafter\ifx\csname url\endcsname\relax
  \def\url#1{\texttt{#1}}\fi
\expandafter\ifx\csname urlprefix\endcsname\relax\def\urlprefix{URL }\fi
\expandafter\ifx\csname href\endcsname\relax
  \def\href#1#2{#2} \def\path#1{#1}\fi

\bibitem{bond1986project}
A.~Bond, A.~R. Martin, Project daedalus reviewed, Journal of the British
  interplanetary society 39~(9) (1986) 385--390.

\bibitem{long2010project}
K.~Long, R.~Obousy, A.~Tziolas, A.~Mann, R.~Osborne, A.~Presby, M.~Fogg,
  Project icarus: Son of daedalus, flying closer to another star, arXiv
  preprint arXiv:1005.3833 (2010).

\bibitem{parkin2018breakthrough}
K.~L. Parkin, The breakthrough starshot system model, Acta astronautica 152
  (2018) 370--384.

\bibitem{RefNumber1015}
K.~Parkin, A starshot communication downlink, arXiv preprint arXiv:2005.08940
  (October 2019).

\bibitem{RefNumber833}
D.~G. Messerschmitt, P.~Lubin, I.~Morrison, Challenges in scientific data
  communication from low-mass interstellar probes, The Astrophysical Journal
  Supplement Series 249~(2) (2020) 36.

\bibitem{messer3}
D.~Messerschmitt, P.~Lubin, I.~Morrison, Technological challenges in low-mass
  interstellar probe communication, in: Tennessee Valley Interstellar
  Symposium, Wichita, KN, 2019.

\bibitem{RefNumber811}
P.~Lubin, A roadmap to interstellar flight, Journal of the British
  Interplanetary Society 69 (2016).
\newblock \href {http://arxiv.org/abs/arXiv:1604.01356}
  {\path{arXiv:arXiv:1604.01356}}.

\bibitem{RefNumber1030}
P.~Lubin, W.~Hettel, \href{https://doi.org/10.5281/zenodo.3874099}{The path to
  interstellar flight}, Acta Futura 12 (2020) 9--44.
\newblock \href {https://doi.org/10.5281/zenodo.3874099}
  {\path{doi:10.5281/zenodo.3874099}}.
\newline\urlprefix\url{https://doi.org/10.5281/zenodo.3874099}

\bibitem{merton1972analytic}
R.~Merton, An analytic derivation of the efficient portfolio frontier, Journal
  of financial and quantitative analysis 7~(4) (1972) 1851--1872.

\bibitem{jakob2014pareto}
W.~Jakob, C.~Blume, Pareto optimization or cascaded weighted sum: A comparison
  of concepts, Algorithms 7~(1) (2014) 166--185.

\bibitem{RefNumber851}
N.~Kulkarni, P.~Lubin, Q.~Zhang, Relativistic spacecraft propelled by directed
  energy, The Astronomical Journal 155~(4) (2018) 155.
\newblock \href {https://doi.org/10.3847/1538-3881/aaafd2}
  {\path{doi:10.3847/1538-3881/aaafd2}}.

\end{thebibliography}
\end{document}